\documentclass[%
 aip,
amsmath,amssymb,
 reprint,%
]{revtex4-1}

\usepackage{graphicx}
\usepackage{dcolumn}
\usepackage{bm}
\usepackage{tabularx}
\usepackage[utf8]{inputenc}
\usepackage[T1]{fontenc}
\usepackage{mathptmx}
\usepackage{etoolbox}
\usepackage[caption=false,font=normalsize,labelfont=sf,textfont=sf]{subfig}
\usepackage{floatrow}
\usepackage{amsmath,amsfonts}
\usepackage{ upgreek }
\usepackage[export]{adjustbox}
\makeatletter
\def\@email#1#2{%
 \endgroup
 \patchcmd{\titleblock@produce}
  {\frontmatter@RRAPformat}
  {\frontmatter@RRAPformat{\produce@RRAP{*#1\href{mailto:#2}{#2}}}\frontmatter@RRAPformat}
  {}{}
}%
\makeatother

\begin{document}

\preprint{AIP/123-QED}

\title{Gap Engineered Superconducting  Multilayer Nanobridge Josephson Junctions}
\affiliation{ 
James Watt School of Engineering, University of Glasgow 
}%
\author{Giuseppe Colletta}


\author{Susan Johny}%
\author{Hua Feng}%
\author{Mohammed Alkhalidi}%
\author{Jonathan A. Collins}
\author{Martin Weides}


\date{\today}

\begin{abstract}
We report the realization of multilayer three-dimensional nanobridge Josephson junctions based on Nb/NbN and Nb/TiN superconducting stacks fabricated using electron beam lithography and chlorine based dry etching. In this architecture, a high resistivity nitride layer defines the geometrical weak link, while the top Nb layer sets the overall critical temperature and film quality of the stack. This multilayer design enables engineering of the superconducting gap and proximity effects without relying on focused ion beam milling or oxide tunnel barriers. The devices are successfully integrated into dc SQUIDs, demonstrating reliable circuit level operation. By combining material selectivity with three-dimensional geometry, this platform provides a scalable route toward oxide free Josephson junctions suitable for superconducting electronics.
\end{abstract}

\maketitle

Nanobridge based Josephson junctions have emerged as an attractive platform for superconducting quantum circuits due to their compact geometry, high critical current density, and absence of lossy tunnel barriers. Their negligible parasitic capacitance and compatibility with single layer or quasi-single layer fabrication processes make them well suited for scalable superconducting electronics and sensitive SQUID architectures \cite{Faley_2021, mypaper, MiniaturizationJJ, RevLuk79, RevGol2004, intrtrasmon, devoret2004superconducting, hertzberg2021laser, collins, Faley_2021, TiNnanobridge, germansquidres, nplnanoSQUIDres, 2023_Kup_Gol, Vijayjunction}. As superconducting circuits progress toward higher integration densities and reduced feature sizes, the scaling of conventional superconductor-insulator-superconductor tunnel junctions becomes increasingly challenging \cite{Burnett2019, PhysRevLettMartin}. In particular, junction miniaturization is constrained by parameter uniformity and by the large intrinsic capacitance of tunnel barriers, which often requires additional circuit elements to control hysteresis \cite{MiniaturizationJJ}. These considerations motivate the exploration of barrier free weak-link Josephson junctions. In nanobridge Josephson junctions, the weak link is defined geometrically as a narrow superconducting constriction. This approach combines intrinsic self shunting behavior with a compact footprint and reduced parasitic capacitance, making it particularly attractive for nanoscale superconducting circuits. Early work on aluminum nanobridges demonstrated that three-dimensional constrictions, in which the weak link is thinner than the adjacent superconducting banks, produce current phase relations close to the short weak link limit and enhanced critical current modulation compared to planar geometries \cite{VijaySquid, Vijayjunction}. In these devices, the thicker banks act as phase reservoirs that confine the order parameter variation to the bridge region and enable more ideal current phase relationships (CPRs). More recently, nanobridge Josephson junctions based on higher critical temperature superconductors such as niobium have enabled operation at elevated magnetic fields and temperatures \cite{germansquidres} showing histeresys free IV curves \cite{collins}. Microwave circuits incorporating 3D nanobridge SQUIDs have demonstrated strong flux responsivity and tunability, highlighting their suitability for a broad range of superconducting circuit applications \cite{germansquidres, nplnanoSQUIDres}. In this work, we extend nanobridge Josephson technology by realizing multilayer Nb/NbN and Nb/TiN 3D nanobridge junctions in which the bottom nitride layer forms the weak link. The devices are fabricated using high resolution electron beam lithography and dry etching, enabling reproducible definition of nanoscale constrictions. We further characterize dc SQUIDs incorporating these multilayer nanobridges, establishing their suitability for circuit level integration.

\begin{figure*}[t]
    \centering
    \includegraphics[width=\linewidth]{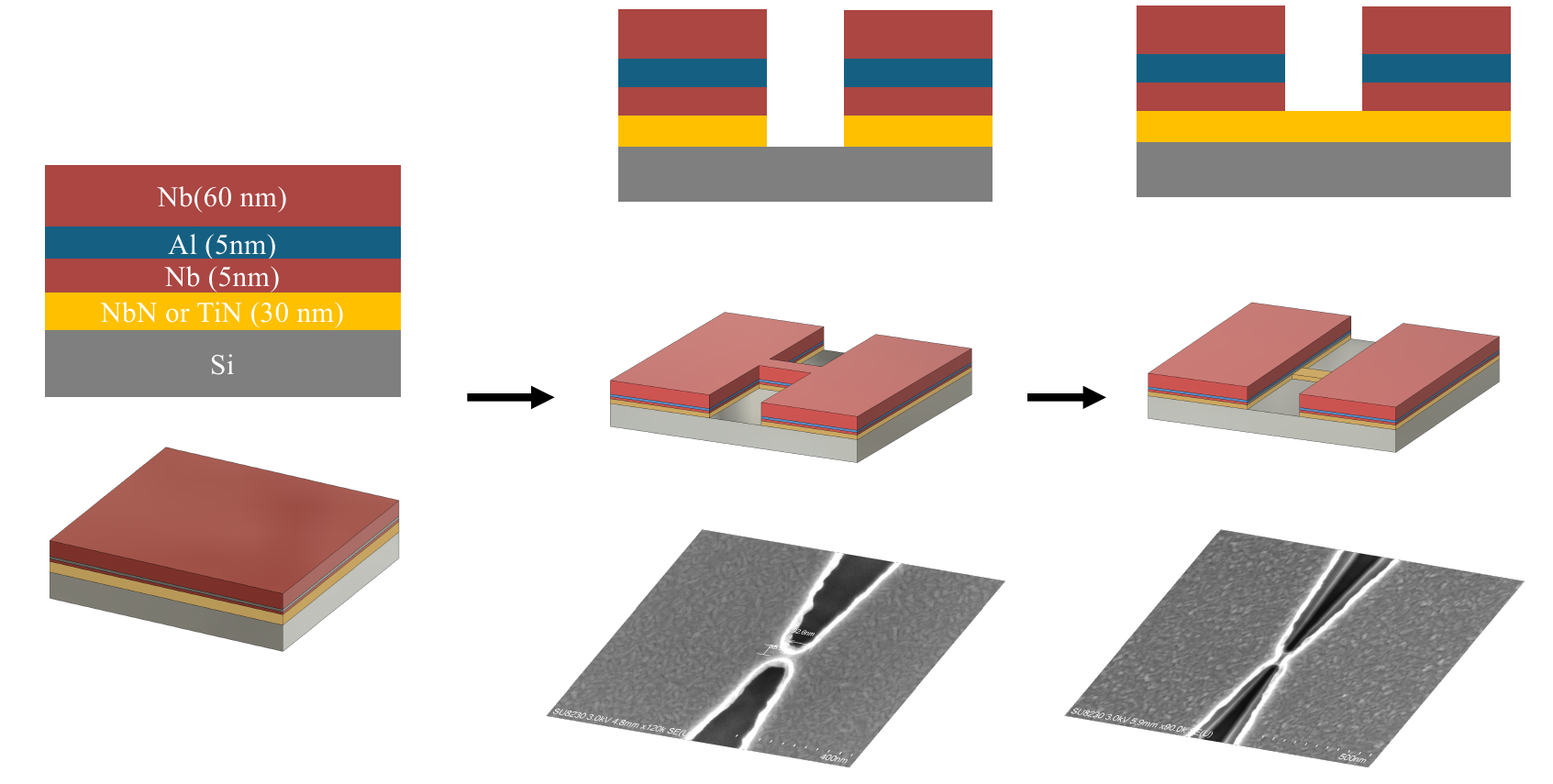}
    \caption{Schematic overview of the two step fabrication process used to define multilayer 3D nanobridges. The process begins with etching the large features and the planar nanobridge structure. In the second patterning step, only the top portion of the film above the nanobridge is etched, leaving the bottom nitride layer intact to form a high resistance 3D weak link geometry.}
    \label{fig:fabrication}
\end{figure*}

All the devices were fabricated on a silicon wafer of thickness 525 $\upmu \mathrm{m}$, prepared by first cleaning 5 min in acetone ultrasonic bath followed by other 5 minutes in IPA after drying with a nitrogen gun, the native oxide is removed from the substrate by submerging it for 2 minutes in a 1:10 HF solution, dried and finally loaded in the chamber for sputtering. Two multilayer superconducting stacks were investigated: Sample A: NbN (50~nm) / Nb (5~nm) / Al (5~nm) / Nb (60~nm), Sample B: TiN (35~nm) / Nb (5~nm) / Al (5~nm) / Nb (60~nm). All layers were deposited in situ at room temperature using dc magnetron sputtering with a base pressure below $10^{-8}$~Torr \cite{propertiesTiN, TIN_variable_Temp, PropertiesNb, Alproperties}. Deposition began with the bottom nitride layer, followed by thin Nb and Al interlayers and a top Nb film. Deposition parameters, summarized in Table \ref{tab:tabsputter}, where found by following previous optimization in the same tool presented in other works of our group \cite{collins,paniz1, paniz2, Poorgholam-shima}. The multilayer approach ensures sharp interfaces and no oxidation between consecutive layers. The thin Al interlayer serves as an optical and interferometric etch stop, facilitating reliable identification of the interface between the top Nb film and the underlying nitride. A thin Nb spacer layer is included to prevent direct contact between the nitride and aluminum, preserving metallic continuity and interface transparency which could be disturbed by the formation of AlN compounds known to be insulating\cite{AlN_1,AlN_2}.

\setlength{\tabcolsep}{0.3em}
\begin{table}[h]
\centering
{\renewcommand{\arraystretch}{1.4}
\begin{tabular}{c c c c c c}
\hline
\hline
Material & Gas [sccm] & Pressure [mTorr] & I [A] & V [V] & P [W]\\
\hline
Nb & Ar: 45 & 3.72 & 0.9 & 260 & 230\\
NbN & Ar: 30, $N_2$: 2.6 & 3.67 & 0.85 & 268 & 227\\
TiN & Ar: 30, $N_2$: 3.5 & 4.08 & 0.85 & 304 & 259\\
Al & Ar: 24 & 2.91 & 0.4 & 350 & 140\\
\hline\hline
\end{tabular}}
\caption{Sputtering deposition parameters for all materials optimized following previous work in our group with the same tool.}
\label{tab:tabsputter}
\end{table}

To define the multilayer structures, we employed a two step chlorine based dry-etching process optimized for selectivity between Nb, NbN/TiN, and the underlying Si substrate, see Fig.\ref{fig:fabrication}. All etching steps were performed in an ICP-RIE system using BCl$_3$/Cl$_2$/Ar chemistry, which ensures smooth sidewalls and minimizes redeposition. The first lithographic step defines the planar geometry of the nanobridge and surrounding leads. After fully etching through the multilayer stack, a second step (resist coating, exposure, and dry etching) removes only the top Nb-Al-Nb layers in the junction region, thereby revealing the nitride underlayer which acts as the actual weak link. This approach produces an intrinsic 3D nanobridge with improved scalability compared to FIB milling techniques \cite{nplnanoSQUIDres, fibdamage1, fibdamge2, fibdamage3}. 
At room temperature, resistance measurements of the nanobridges show a $\sim19\%$ spread across nominally identical devices (see Fig. \ref{fig:RT_mes} in the Supplementary Material for additional characterized samples), reflecting a moderate level of variability. 
However, at base temperature we observe a larger spread in the critical current and normal state resistance. This behavior is likely due to the strong sensitivity of nanobridge junctions to small geometric variations \cite{mypaper}, as well as possible localized contamination sources.

\begin{figure*}[t]
\centering
\subfloat[]
{\includegraphics[width=2.6in]{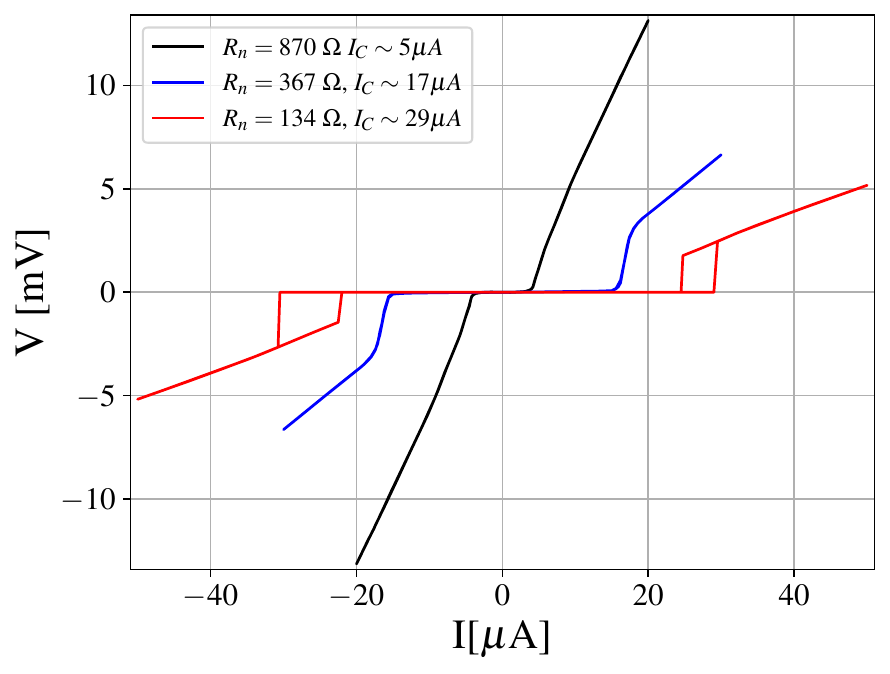}}
\subfloat[]
{\includegraphics[width=2.67in]{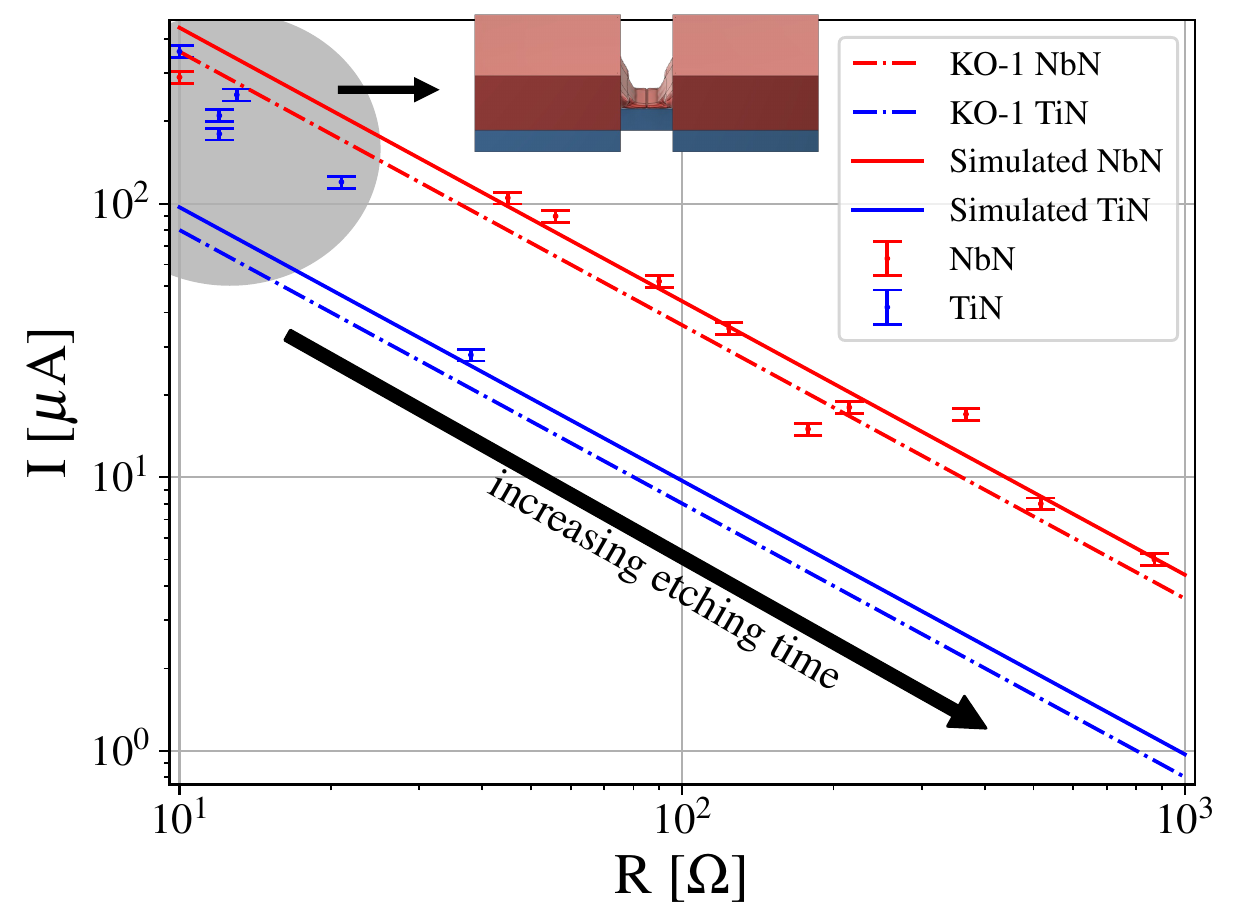}}\quad
\subfloat[]
{\includegraphics[width=1.1in]{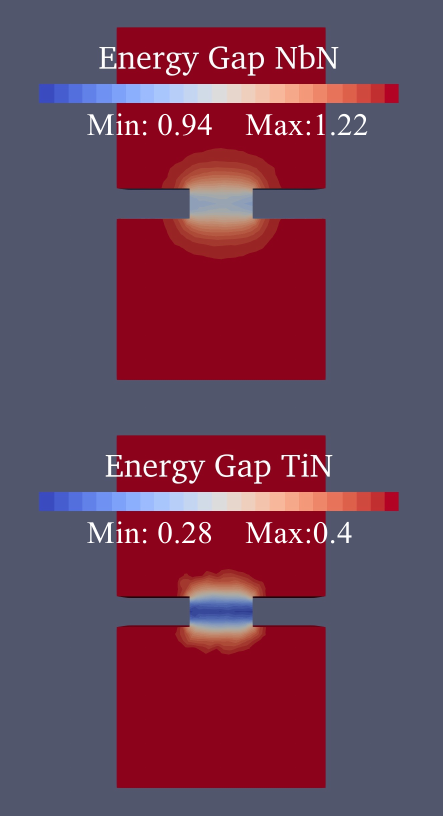}}
\caption{(a) Representative current–voltage (IV) characteristics of NbN based nanobridge Josephson junctions measured at 40~mK. (b) Critical current as a function of normal-state resistance for all measured NbN junctions. Two distinct device populations are observed. Junctions in the gray region exhibit relatively low resistance and are attributed to under etching, where the top Nb layer is not fully removed (schematically shown in the inset). In this regime, the $I_cR_N$ product is independent of the underlying nitride layer due to the residual metallic Nb contribution. For longer etching times, the devices follow the theoretically predicted trend, indicating proper removal of the top layer and correct definition of the nanobridge. The lines correspond to numerical simulations performed using the KO-1 model and our three-dimensional Usadel based model, assuming a bridge length $L=30$~nm and $\xi = 10$ nm for both TiN and NbN as explained in the main text, and adjusting the effective cross section to reproduce the measured normal-state resistance. The level of agreement between simulations and experimental data was quantified through a $\chi_{norm}^2$ analysis. For the NbN junctions $\chi_{norm}^2 = 0.96$ within our model and $3.5$ within the KO-1 framework. (c) Simulated spatial profile of the superconducting energy gap for NbN and TiN based multilayer nanobridge junctions at $T = 0.1T_c$, with $L = 30$~nm and $W = 60$~nm. The structure is shown from the bottom of the junction. Due to proximity effects within the multilayer stack, the superconducting gap in NbN is reduced by approximately $5\%$, whereas in TiN it is enhanced by approximately $60\%$. The smaller variation observed in NbN is attributed to its larger thickness and to a critical temperature closer to that of Nb, which reduces the strength of proximity suppression compared to the TiN case.}
\label{fig:IV curves NbN}
\end{figure*}

Nanobridges from samples A and B were characterized in a dilution refrigerator at 40~mK. Representative current–voltage characteristics for NbN based devices are shown in Fig.~\ref{fig:IV curves NbN}.(a). The junctions exhibit a switching transition to a resistive state consistent Josephson behavior, with $I_cR_N$ products in the range of 1–8~mV. Such values are compatible with superconducting technologies including rapid single flux quantum (RSFQ) circuits, SQUIDs, and flux tunable resonators \cite{RSFQ, Shelly_2017, nplnanoSQUIDres, nanosquid, germansquidres}. To benchmark device performance, we extracted the critical current $I_c$ and normal-state resistance $R_N$ from a set of nominally identical junctions fabricated under the same conditions, while systematically increasing the etch time in the final fabrication step. The resulting $I_cR_N$ distribution is shown in Fig.~\ref{fig:IV curves NbN}.(b), where two distinct device populations can be identified. Properly fabricated junctions follow the trends predicted by both the KO-1 model and our three-dimensional Usadel based simulations. In contrast, under-etched devices (gray region) exhibit lower resistance due to incomplete removal of the top Nb layer; in this regime, both TiN and NbN based junctions display similar characteristics, consistent with a residual metallic Nb contribution dominating transport. The numerical simulations were performed assuming a bridge length $L=30$~nm and $\xi= 10$ nm for the nitride layers, while the effective cross section was adjusted to reproduce the measured normal-state resistance, without introducing additional free parameters. The simulations reveal the redistribution of the superconducting order parameter across the multilayer stack, highlighting the role of proximity effects between the nitride weak link and the Nb layer. These results provide insight into the material dependent trends observed in the critical current measurements. The agreement between experiment and theory was quantified through a $\chi_{norm}^2$ distance analysis:
\begin{equation}
    \chi_{norm}^2=\sum_{i=1}^N \frac{(O_i-E_i)^2}{N E_i}
\end{equation}
where $O_i$ is the measured critical current and $E_i$ is the value obtained either with the KO-1 model or with our numerical approach. For the NbN junctions, which provided the largest dataset, $\chi_{norm}^2 = 0.96$ was obtained for the three-dimensional model, compared to $3.5$ within the KO-1 framework. The reduced $\chi_{norm}^2$ demonstrates that the three-dimensional model captures the experimental trends with good accuracy and highlights the importance of explicitly accounting for multilayer geometry and proximity effects in describing current transport through nanobridges.
The simulated energy gap profiles, shown in Fig.~\ref{fig:IV curves NbN}(c), further clarify the role of proximity effects in the multilayer architecture. In the NbN based junctions, the superconducting gap is reduced by approximately $5\%$ due to proximity coupling with the Nb layer. In contrast, the TiN based junctions exhibit an enhancement of the gap of approximately $60\%$, indicating a stronger redistribution within the stack. These results highlight the ability of the multilayer geometry to engineer the effective superconducting energy scales of the weak link, directly impacting the critical current and transport properties.

\begin{figure}[!t]
\centering
\subfloat[]
{\includegraphics[width=2.2in]{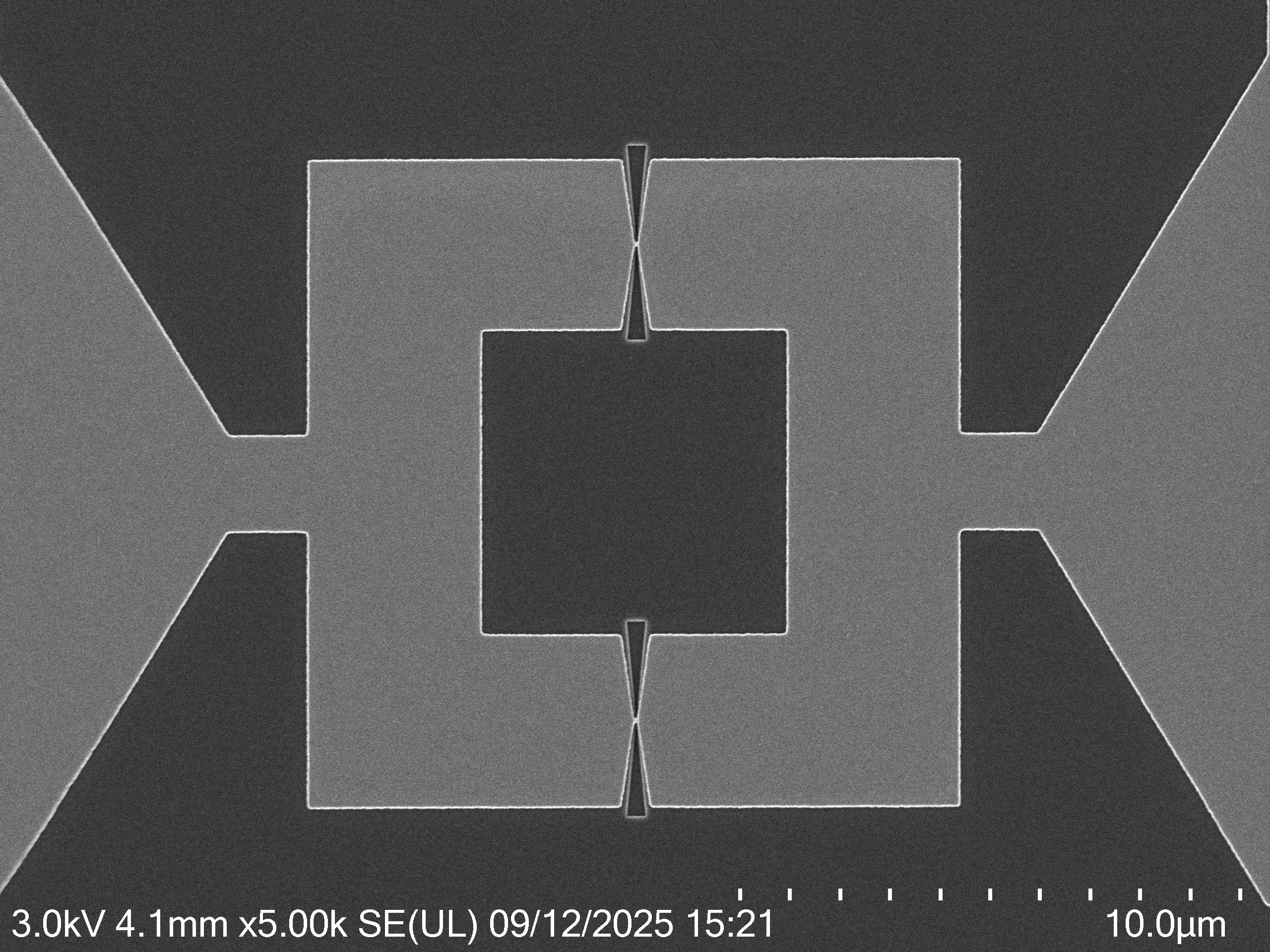}} \quad 
\subfloat[]
{\includegraphics[width=3in]{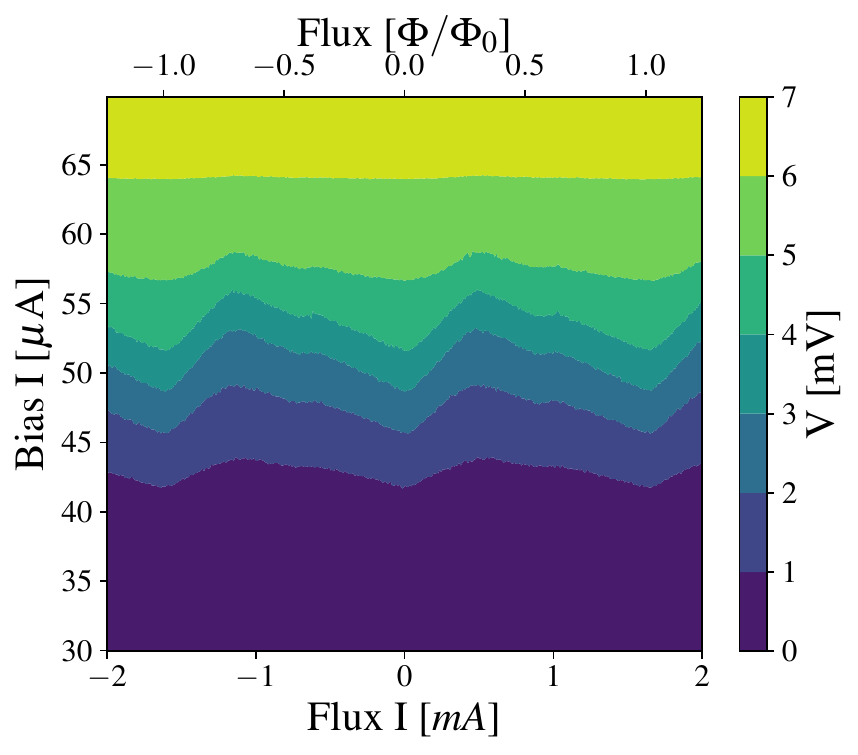}} \quad 
\subfloat[]
{\includegraphics[width=2.92in]{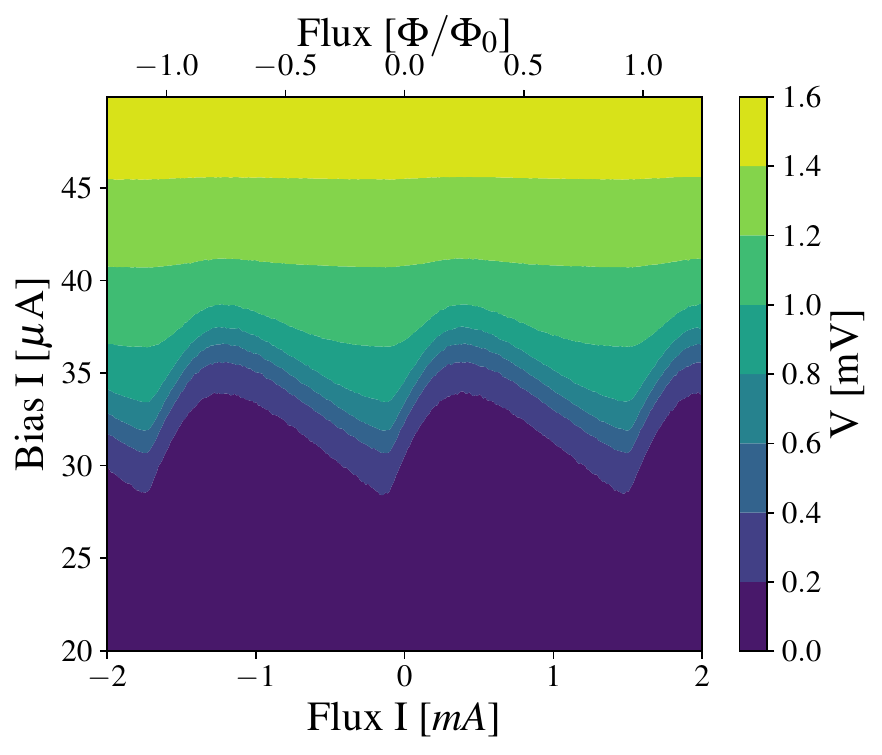}} \quad
\caption{(a) SEM image of a fabricated dc SQUID. Color plots of IV characteristics versus applied flux for (b) NbN based and (c) TiN based devices. The NbN SQUID shows approximately $5\%$ modulation depth, while the TiN device exhibits approximately $20\%$ modulation. The relatively modest visibility is attributed to the large $4 \times 4~\mu m^2$ loop area, resulting in loop inductance comparable to the Josephson inductance, and possible asymmetry between the two junctions.}
\label{fig:SquidIV}
\end{figure}

\begin{figure}[!ht]
\centering
\subfloat[]
{\includegraphics[width=3.5in]{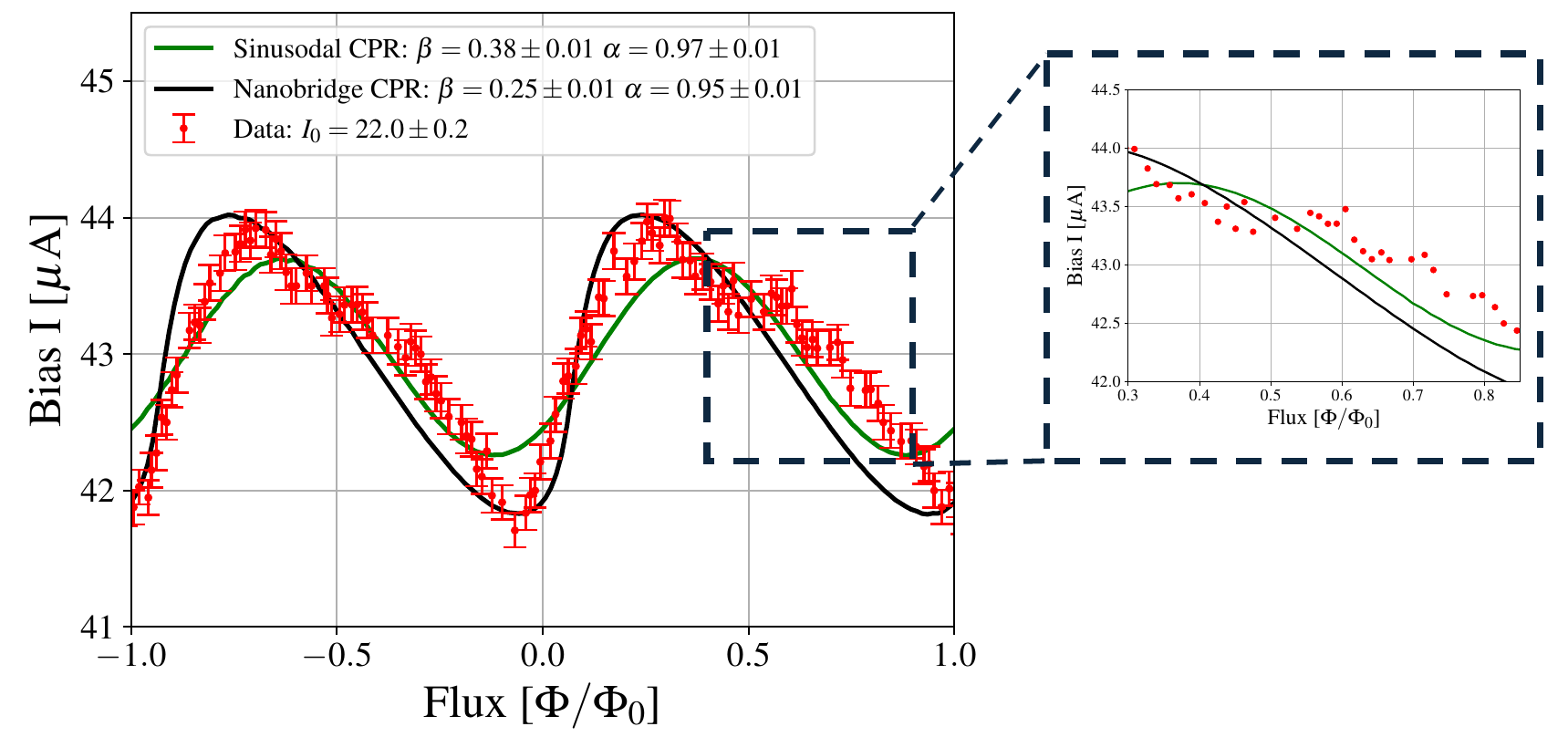}} \quad 
\subfloat[]
{\includegraphics[width=2.92in]{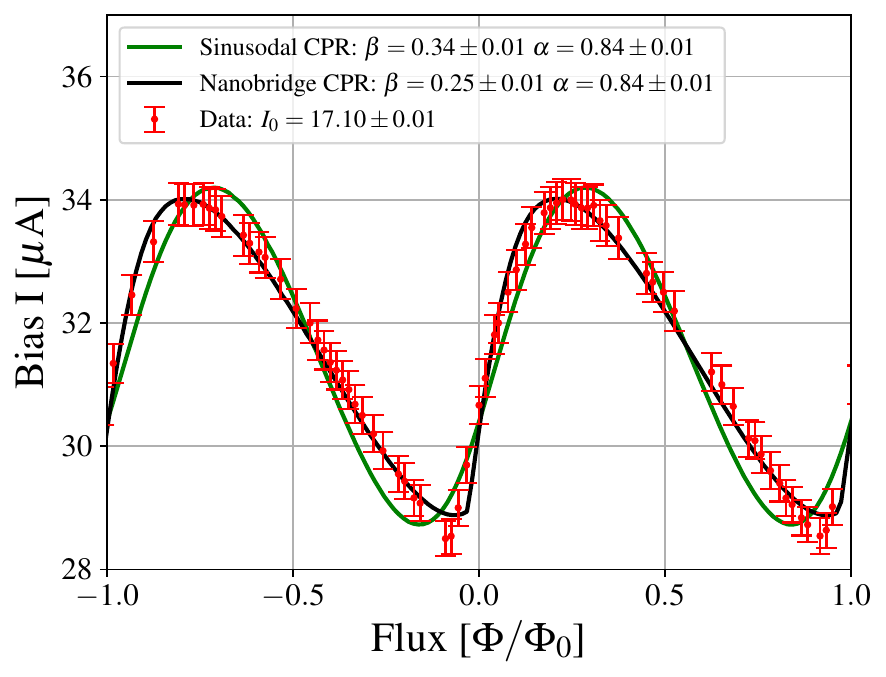}} \quad
\caption{Measured SQUID critical current vs flux and corresponding fits. (a) NbN based sample. The inset shows a zoomed view of the data highlighting a small bump, a typical signature of higher harmonic contributions to the CPR. (b) TiN based sample.
Solid lines show the best fits to the data, accounting for junction asymmetry and
finite loop inductance. The fits were performed using the asymmetric dc SQUID
formalism described in the main text, comparing a sinusoidal current phase relation and the nanobridge current phase relation reported in
Ref.\onlinecite{mypaper}}
\label{fig:SquidIV_fit}
\end{figure}
SQUIDs  were fabricated by placing two nanobridges in a loop of lateral size approximately $4\times 4~\mu$m$^2$.
The micron scale SQUID loop area was deliberately chosen instead of a nano SQUID geometry to allow magnetic flux control via an integrated on-chip flux line rather than an externally applied magnetic field. This design choice enables device characterization under experimentally accessible operating conditions while maintaining sufficient flux coupling. Although significantly larger than nanoscale loops used in state of the art nanoSQUIDs \cite{nplnanoSQUIDres, Faley_2021,germansquidres, TiNnanobridge,nanosquid}, the devices exhibited clear flux modulation of the critical current of approximately 20\% for TiN based junctions and approximately 5\% for NbN based devices (with higher noise levels). The lower modulation depth in NbN devices may be attributed to variations in etching or difference in thickness on the sample locally; other NbN nanobridges on the same chip exhibited lower critical currents compared to the approximately 50~$\mu$A for the SQUID junctions. The experimental SQUID interference patterns were analyzed by fitting the data with
an asymmetric dc SQUID model. In this framework, the normalized total current \(i\)
and the circulating current \(j\) are given by \cite{Squidbook}
\begin{align}
  i &= (1-\alpha)I(\pi \delta_1) + (1+\alpha)I(\pi \delta_2), \\
  j &= (1-\alpha)I(\pi \delta_1) - (1+\alpha)I(\pi \delta_2),
\end{align}
where \(\alpha\) parametrizes the asymmetry between the two Josephson junctions,
\(\delta_1\) and \(\delta_2\) are the superconducting phase differences across each
junction and $I(\delta)$ is the CPR of the junction. The phase constraint imposed by flux quantization reads
\begin{equation}
  \delta_2 - \delta_1 = 2\pi\Phi_{\mathrm{ext}} + \pi\beta_L\, j,
\end{equation}
with \(\Phi_{\mathrm{ext}}\) the externally applied magnetic flux normalized to the
flux quantum. Both \(i\) and \(j\) are normalized to the average critical current
\(I_0\).
the SQUID screening parameter \(\beta_L\), which
quantifies the strength of magnetic flux screening in the loop, is given by
\begin{equation}
\label{eq:beta}
  \beta_L = \frac{2L_{\mathrm{loop}}I_0}{ \Phi_0},
\end{equation}
where \(L_{\mathrm{loop}}\) is the geometric loop inductance and \(I_0\)is the average critical current for the 2 junctions in the squid and \(\Phi_0\) is the flux quantum. The extracted values of $\beta_L$ are found to depend on the assumed current phase relation (CPR). In particular, fitting the data for the TiN based SQUID (sample B) with a purely sinusoidal CPR yields $\beta_L \simeq 0.34 \pm 0.01$. Using instead the more appropriate CPR for nanobridge junctions, as described in our previous work \cite{mypaper}, results in a reduced value $\beta_L = 0.25 \pm 0.01$, together with a critical current $I_0 = 17.10 \pm 0.01~\mu\text{A}$. This value is consistent with an independent estimate of $\beta_L$ obtained from $I_0$, assuming a square loop geometry of area $4 \times 4~\mu\text{m}^2$. For this geometry, the loop inductance is $L_{\mathrm{loop}} \simeq 15~\text{pH}$, yielding $\beta_L \simeq 0.25$ from Eq.~\ref{eq:beta}. A similar behavior is observed for the NbN based SQUID. In this case, fitting the data with a sinusoidal CPR gives $\beta_L \simeq 0.38$, while using the nanobridge CPR leads to a slightly reduced value of $\beta_L \simeq 0.25$. However, both values show a poor agreement with the estimate $\beta_L \simeq 0.32$ obtained from Eq.~\ref{eq:beta}. This discrepancy is likely due to the fact that neither the sinusoidal nor the skewed CPR is sufficient to fully capture the behavior of the NbN nanobridge junctions. Due to the short superconducting coherence length of NbN, higher harmonics can emerge from a multivalued current–phase relation, as shown in previous theoretical and experimental works \cite{Giazzotto3dsquid, diffcprth, diffcprexp}. Consistently with this interpretation, a small bump is observed in the modulation curve around $0.5\,\Phi_0$ in Fig.\ref{fig:SquidIV_fit}.(a), which is a typical signature of higher harmonic contributions to the CPR \cite{diffcprth}. An additional contribution may originate from asymmetries between the CPRs of the two nanobridges forming the SQUID, which can further suppress the modulation depth. At present, however, the coexistence of multivalued CPR effects, higher harmonics, and junction asymmetry introduces too many competing factors to be accurately captured. Moreover, both the TiN and NbN based SQUIDs exhibit a pronounced junction asymmetry, with $\alpha > 0.8$. Such a high degree of asymmetry is likely related to the nanobridge dimensions being comparable to the superconducting coherence length and possible contaminations sources, which makes the critical current highly sensitive to small fabrication induced variations. This sensitivity may also lead to the aforementioned asymmetries in the CPRs of the two junctions forming the SQUID.

We have demonstrated a multilayer nanobridge fabrication process based entirely on electron beam lithography and chlorine based dry etching, enabling the realization of variable thickness nanobridge (VTB) Josephson junctions without the use of focused ion beam milling. Using this approach, we successfully fabricated Nb/NbN and Nb/TiN nanobridges that exhibit well defined switching behavior and clear SQUID interference patterns, confirming that Josephson junctions and functional SQUIDs can be reliably realized with this process. While the measured device characteristics show a strong dependence on the geometric dimensions of the nanobridge, which are currently difficult to control with high precision in the present fabrication flow, the overall results are encouraging. The observation of reproducible Josephson behavior across multiple devices demonstrates the robustness of the multilayer approach and validates the feasibility of nanobridge based junctions fabricated using standard nanofabrication techniques. The use of multilayer superconducting stacks provides additional degrees of freedom for engineering junction properties, including the critical current and the current–phase relation, which are not easily accessible in conventional tunnel junctions. Future work will focus on improving dimensional control in the second etching step, optimizing etch selectivity between layers, and reducing SQUID loop dimensions to enhance flux modulation depth and device uniformity. These improvements will be essential for transitioning nanobridge Josephson junctions to scalable elements for superconducting quantum circuits.

\section*{Acknowledgements}
This work was supported by the UK Engineering and Physical Sciences Research Council (EPSRC) grant EPIT025746/1 for the University of Glasgow. The authors acknowledge the James Watt Nanofabrication Centre for device fabrication and thank the staff and in particular Paul Reynolds for their support and valuable discussions in defining the process. We also thank Alessandro Casburi and Valentino Seferai for helpful discussions.

\section*{Conflict of Interest}
The authors have no conflicts of interest to disclose.
\subsection*{ AUTHOR CONTRIBUTIONS}
\textbf{Giuseppe Colletta}: Conceptualization (equal), Software (lead), Methodology (lead), Investigation (lead), Writing/Original Draft Preparation (lead),  Writing – review and editing (equal). \textbf{Susan Johny}:  Conceptualization (equal), Investigation (equal), Writing – review and editing (equal). 
\textbf{Hua Feng}: Investigation (supporting),
\textbf{Mohammed Alkhalidi}: Investigation (supporting),\textbf{Jonathan A. Collins}: Conceptualization (equal), Writing – review and editing (equal). \textbf{Martin Weides}: Conceptualization (equal), Writing – review and editing (equal), Funding Acquisition (lead). 
\section*{Data Availability}
The data that support the findings of this study are available from the corresponding author upon reasonable request.
\section*{References}
\bibliography{aipsamp}
\clearpage
\section{Supplementary Material}
\renewcommand{\theequation}{S.\arabic{equation}}
\renewcommand{\thefigure}{S.\arabic{figure}}
\setcounter{equation}{0}
\setcounter{figure}{0}
\begin{figure}[!h]
\centering
\subfloat[]
{\includegraphics[width=2.5in]{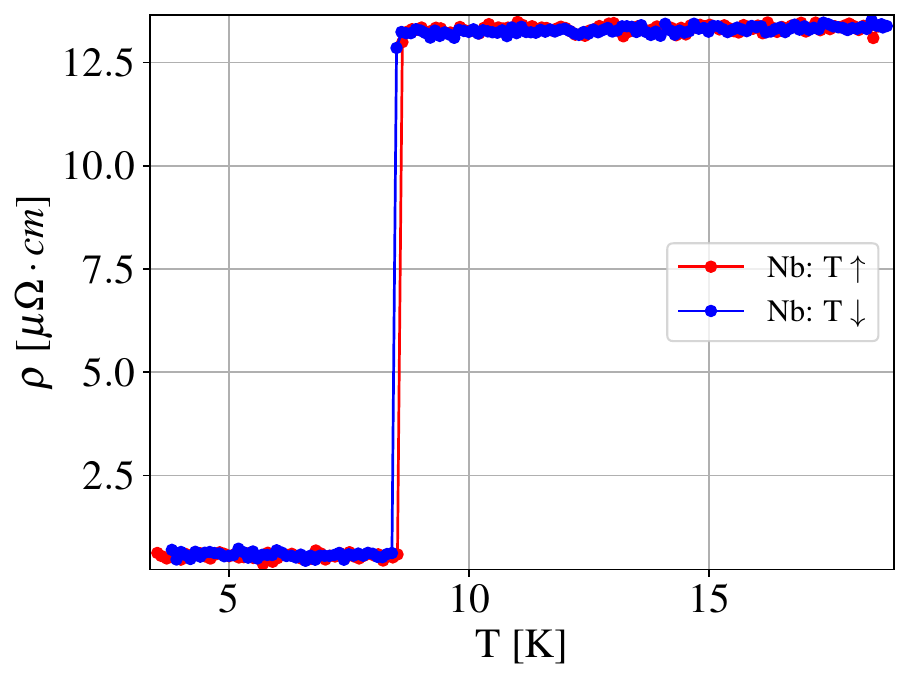} 
}\\
\subfloat
{\includegraphics[width=2.5in]{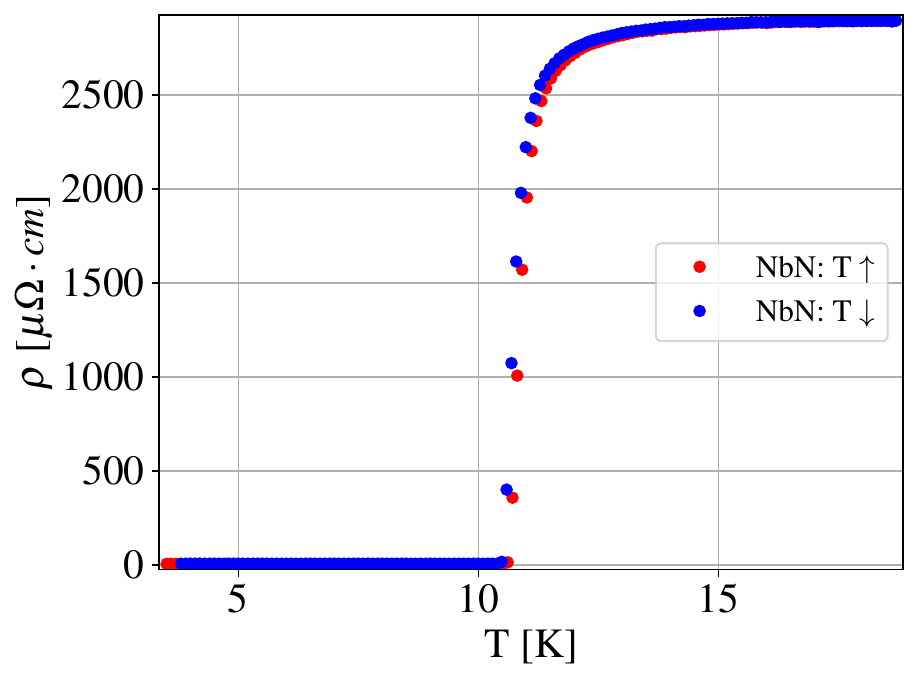}
}\\
\subfloat[]
{\includegraphics[width=2.5in]{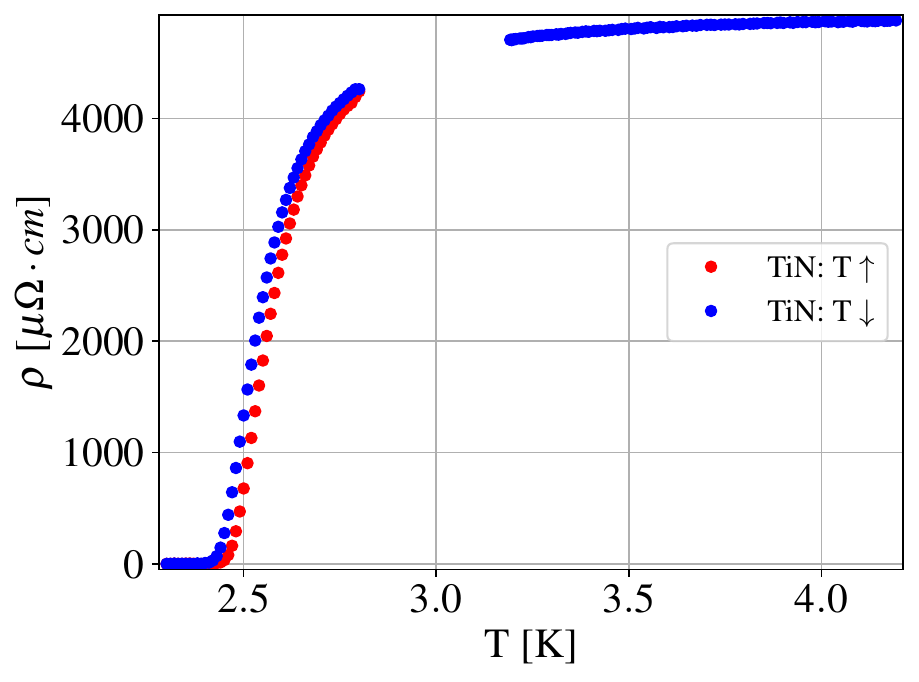}
}
\caption{Critical temperature $T_C$ and resistivity characterization of the materials used for nanobridge fabrication: (a) Nb, (b) NbN, and (c) TiN. The absence of data points around $3\,\mathrm{K}$ for TiN is due to measurements performed in a different cryostat, which did not allow fine temperature control in this range.}
\label{fig:Tcmaterial}
\end{figure}

\subsection{Film Characterization}
\label{sec:film_characterization}

The superconducting films used for nanobridge fabrication (Nb, NbN, and TiN) were deposited and independently characterized by means of four point resistivity measurements in order to determine their critical temperature $T_C$ and normal-state resistivity, see Fig.~\ref{fig:Tcmaterial}. Measurements were performed using a standard four probe technique on patterned test structures consisting of $10$ square wires. This geometry was chosen to improve the accuracy of the resistivity extraction by reducing the influence of contact resistance.  The temperature dependence of the resistance was recorded during slow cooling/warm-up, allowing precise determination of the superconducting transition between the normal and superconducting states. The critical temperature $T_C$ was defined as the midpoint of the resistive transition, while the normal-state resistivity was extracted from the linear region above $T_C$. The extracted superconducting transition temperatures and material parameters are summarized in Table~\ref{tab:tabTC}. The measured residual resistivity ratios are consistent with values typically reported for sputtered thin films \cite{TiNproperties, nbntype2, faileytin}. From the measured resistivity and critical temperature, we estimated the dirty limit coherence length as \cite{tinkham2004introduction, Giazzotto3dsquid}
\begin{equation}
    \xi = \sqrt{\frac{\hbar}{\rho \cdot 1.76 k_B T_C e^2 N_F(E_F)}},
\end{equation}

\setlength{\tabcolsep}{0.3em}
\begin{table}[!b]
\centering
{\renewcommand{\arraystretch}{1.4}
\begin{tabular}{c c c c c c c}
\hline
\hline
Material & $T_C$ [K]& t [nm] &$\rho$ [$\mu\Omega \cdot$cm] & RRR & $E_F$ [eV] & $\xi$ [nm]\\
\hline
Nb & 8.45 & 60  & 13 & 2.79 & 5.3\cite{NbFermi} & 29\\
NbN & 10.55 & 80 & 2890 & 0.75 & 1.7\cite{PropertiesNbN} & 2.2\\
TiN & 2.35 & 60 & 4810 & 0.94 & 2.9\cite{TiNfermi} & 3.5\\
\hline
\hline
\end{tabular}}
\caption{Superconducting film properties.}
\label{tab:tabTC}
\end{table}

Here $N_F(E_F)$ is the density of states at the Fermi level. Literature values were used to approximate the Fermi energy and corresponding density of states for each material. For the nitride films, this approach yields relatively small coherence length values. Given the sensitivity of $\xi$ to uncertainties in $N_F(E_F)$ and to material disorder, the actual coherence length is expected to deviate from this simplified estimate. For this reason, in the numerical analysis we adopt an effective coherence length of $\xi = 10$~nm for the nitrides. This value provides a physically consistent description of the experimental trends discussed in the main text. These independently measured material parameters were subsequently used as input for the numerical modelling of the nanobridge Josephson junctions presented in the main text.

\begin{figure}[!t]
\centering
\includegraphics[width=2.5in]{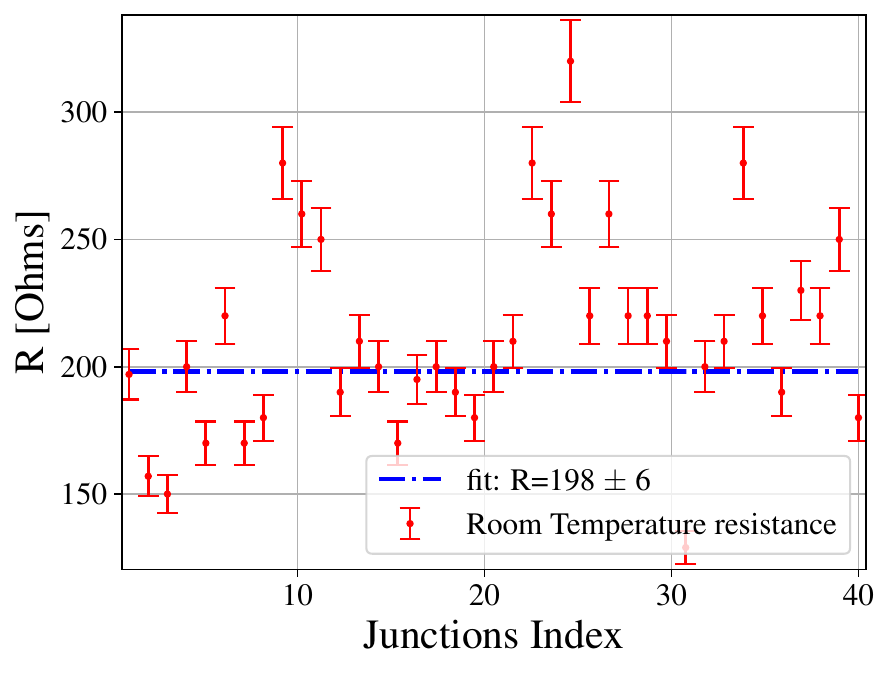}
\caption{Room temperature resistance of multilayer nanobridges measured using a probe station. The resistance values are plotted as a function of the junction index for a set of nominally identical devices. The distribution is centered at $R = 198~\Omega$, with a spread of $\sim 20\%$ across devices, indicating moderate variability likely arising from fabrication induced inhomogeneities. Despite this, the overall resistance scale and trends remain consistent, supporting the reproducibility of the nanofabrication process.}
\label{fig:RT_mes}
\end{figure}

\begin{figure}[!t]
\centering
\subfloat[]
{\includegraphics[width=2.5in]{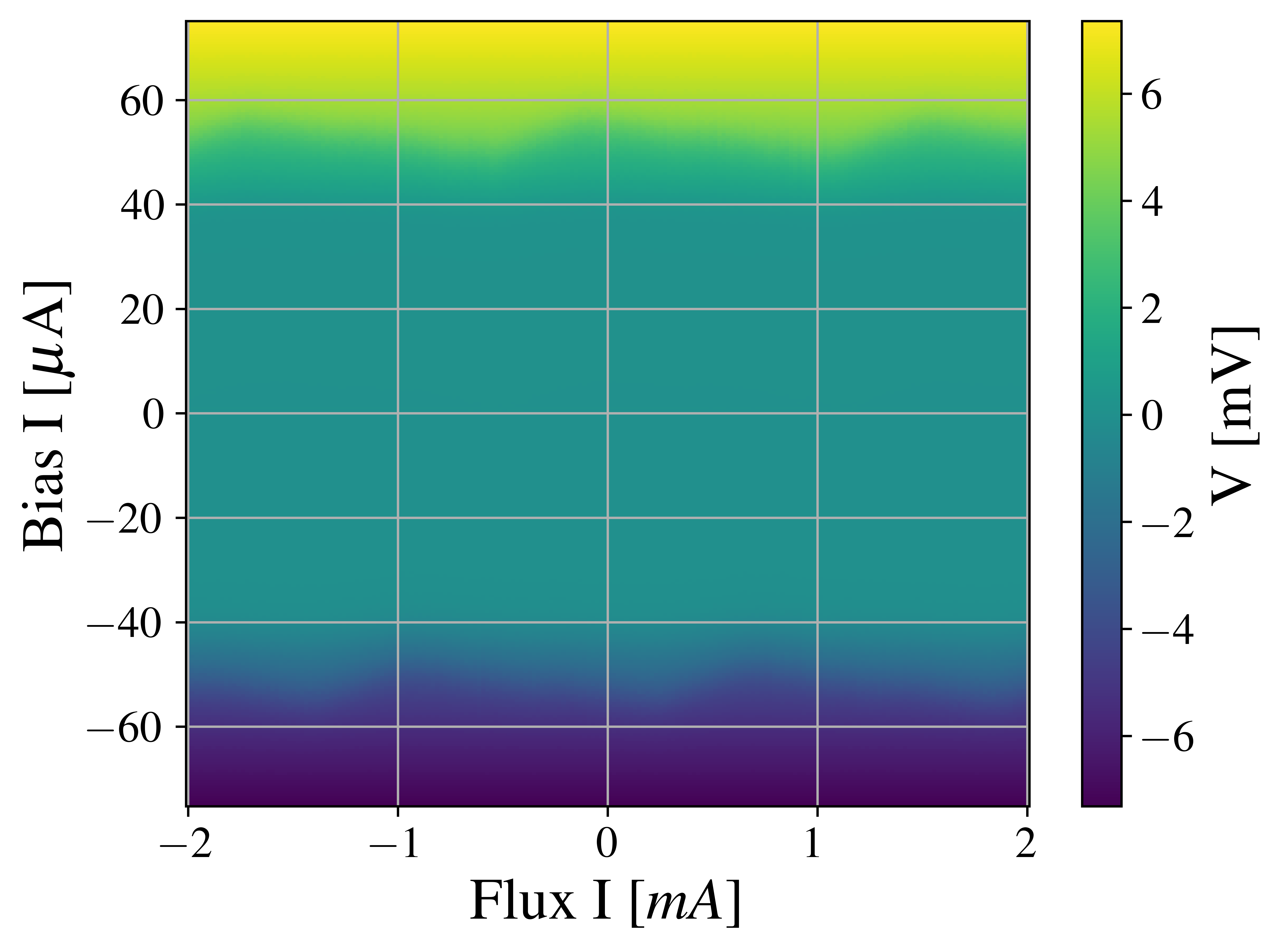}} \\
\subfloat[]
{\includegraphics[width=2.5in]{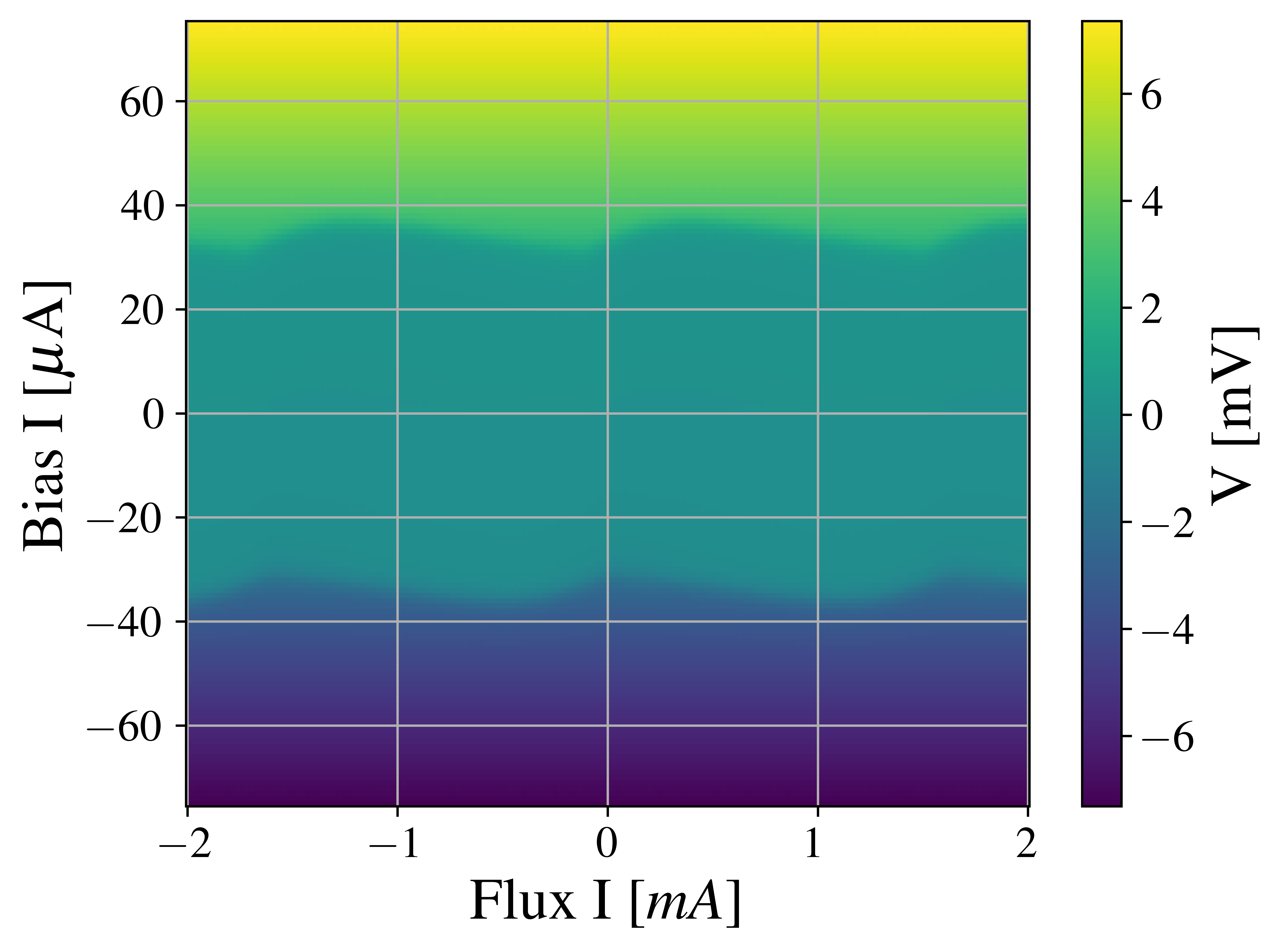}} 
\caption{Sweep of the bias current from negative to positive values, showing the full SQUID
response. The asymmetry between the positive and negative bias branches reflects the
difference in the critical currents of the two junctions.}
\label{fig:SquidIV_full}
\end{figure}
\subsection{Reproducibility of the nanobridge fabrication process}
\label{sec:RT_Nano}
To evaluate the reproducibility of the nanofabrication process, we measured the room temperature resistance of a series of nominally identical multilayer nanobridges using a probe station. Each device was contacted individually and the resistance was recorded under ambient conditions. The measured resistances are shown in Fig.~\ref{fig:RT_mes} as a function of the junction index. The resistance distribution is centered at $R \approx 198~\Omega$, with a spread of $\sim 20\%$ across devices. This level of variability likely reflects fabrication induced inhomogeneities at the nanoscale. Despite this, the overall resistance scale and the absence of systematic outliers indicate consistent device behavior, supporting the reproducibility of the multilayer nanofabrication process.

\subsection{SQUIDs}
\label{sec:squidfull}
For completeness, we report here the full flux modulation measurements of the SQUID devices. Fig.\ref{fig:SquidIV_full} shows the bias current swept continuously from negative to positive values, displaying the complete SQUID response. The full sweep allows visualization of both positive and negative critical current branches. A clear asymmetry between the two branches is observed, manifested as a shift for opposite bias polarities. This behavior reflects the difference in the critical currents of the two nanobridge junctions forming the SQUID loop, which is expected given the sensitivity of nanobridge properties to small geometric variations introduced during fabrication.
\clearpage
\end{document}